\documentclass{elsart}
\usepackage{amsmath}
\usepackage{amssymb}
\usepackage[a4paper]{geometry}
\usepackage{graphicx}
\usepackage{psfrag}
\usepackage{epsfig}
\begin{document}

\def \bsigma {\mbox {\boldmath $\sigma$}}
\begin{frontmatter}

\title{The three-state layered neural network with finite dilution}
\author{W. K. Theumann\corauthref{cor}}
\corauth[cor]{Corresponding author. 
Phone: + 55-51-3316.6486. Fax: + 55-51-3316.7286}
\ead{theumann@if.ufrgs.br}
and
\author{R. Erichsen Jr.}
\ead{rubem@if.ufrgs.br}
\address{Instituto de F{\'\i}sica, Universidade Federal do Rio
Grande do Sul\\
Caixa Postal 15051, 91501-970 Porto Alegre, RS, Brazil}







\begin{abstract}
The dynamics and the stationary states of an exactly solvable
three-state layered feed-forward neural network model with
asymmetric synaptic connections, finite dilution and low pattern
activity are studied in extension of a recent work on a recurrent
network. Detailed phase diagrams are obtained for the stationary
states and for the time evolution of the retrieval overlap with a
single pattern. It is shown that the network develops
instabilities for low thresholds and that there is a gradual
improvement in network performance with increasing threshold up to
an optimal stage. The robustness to synaptic noise is checked and
the effects of dilution and of variable threshold on the
information content of the network are also established.



\end{abstract}
\end{frontmatter}

\section{Introduction}
The statics and the dynamics of large attractor and feed-forward
neural networks, inspired in features of biological networks, have
been studied over some time. Some of these features are the
asymmetry and the finite dilution of the synaptic connections as
well as the low activity of the patterns and the neurons. It has
been suggested that a recurrent attractor neural network model of
multi-state neurons with these characteristics may describe the
short term memorization performance of the $CA_3$ region of the
hyppocampus in both the human brain and in the brain of primates,
and results of numerical simulations on the retrieval behavior are
now available \cite{RTFP97}. Full results on the dynamic evolution
of the retrieval overlap and on the phase diagrams for the
stationary states of low-activity networks of multi-state neurons
with asymmetric interactions and finite dilution are still
missing. It would be desirable to have such results in order to
describe other implementations of neural network models, among
them as devices to account for long-term memory in the brain
\cite{Ar95}.

The retrieval behavior and thermodynamic properties of
symmetrically diluted Q-Ising recurrent neural networks with
finite dilution and low activity patterns have been studied and
full phase diagrams have been obtained for $Q=3$ and $Q=4$ states
as well as for a network with continuous response neurons
\cite{TE01}. In the case of a discrete number of states, phase
boundaries between states of low and high performance are found to
disappear beyond a finite dilution, allowing for the biologically
appealing feature of a continuous improvement of the behavior of
the network without the need of a precise threshold adjustment.
The full study of the dynamics of these networks is rather
involved (see ref. \cite{Co01}) and for practical, either hardware
or biological implementations of neural networks, it would be
interesting to have a simple dynamical system.

A suitably tractable model to study these issues is a feed-forward
layered network with no feedback loops, which has an exactly
solvable non-trivial dynamics, in which the only non-zero synaptic
connections are the asymmetric interactions that pass information
from one layer to the next. Despite the fact that the connectivity
of the layered network is much lower than that of a recurrent
network, the presence of non-trivial correlations makes it an
ideal model to test the qualitative behavior of feature dependence
in recurrent networks. The study of the model in itself is also of
interest in view of the practical applications of feed-forward
layered networks.

The purpose of the present work is to present new results on
the retrieval performance, the information content and the dynamic
evolution for this network with three-state Ising neurons and finite
dilution. Results for the diluted layered network with binary units
and for the three-state network with no dilution can be found in the
literature \cite{DKM89,BSV94}.

The outline of the paper is the following. In Section 2 we
introduce the model and the relevant macroscopic variables. In
Section 3 we derive the recursion relations for these variables
that establish the dynamics for the model. We discuss the results
for the phase diagrams of the stationary states and the dynamical
evolution of the retrieval overlaps in Section 4, and end with
concluding remarks in Section 5.

\section{The Model}

The network model consists of $L$ layers with $N$ neurons on each
layer, that can take values $\sigma_i^{l}$, $l=1,...,L; i=1,...,N$ from
the set ${ S}\equiv\{-1,0,+1\}$, where $\pm 1$ denote the active states.
A macroscopic number of $p=\alpha cN$ ternary patterns, where $c$ is the
probability that two neurons are connected, is taken from a set of
independent identically distributed random variables
$\{\xi^{{\mu},l}_i=0,\pm 1\}$, $\mu=1,...,p$, with the probability
distribution
\begin{equation}
\mbox{Prob}(\xi^{{\mu},l}_i)
= a\delta(|\xi^{{\mu},l}_i|^{2}-1)+(1-a)\delta(\xi^{{\mu},l}_i) \,
\label{1}
\end{equation}
which is assumed to be the same for every layer and where $\pm 1$
are the active patterns. The mean of each pattern (over-lined) is
zero and $a=\overline{(\xi^{{\mu},l}_i)^2}$, denotes their
activity. A new random set of $p$ patterns is generated on each
layer and the whole set on two consecutive layers is embedded in
the diluted network according to the generalized Hebbian learning
rule
\begin{equation}
J_{ij}^{l}=
\frac{c_{ij}^l}{acN}\sum_{\mu=1}^{p} \xi^{{\mu},l+1}_i\xi^{{\mu},l}_j \,\,,
\label{2}
\end{equation}
where $\{c_{ij}^l\}$ is a set of independent identically
distributed random variables that account for the dilution of the
synapses, particularly when the patterns are active, and such that
$c_{ij}^l=1$ with probability $c$ and zero with probability $1-c$,
for all $l$. Thus, $cN$ is the mean number of neurons connected to
each neuron. For the fully connected layered network, $c=1$, and
the case of extreme dilution corresponds to the limit
$c\rightarrow 0$, after taking the macroscopic
$N\rightarrow\infty$ limit. The dilution introduces an additional
randomness into the dynamics of the fully connected network in the
form of a static noise $z_{ij}^l/\sqrt{N}$ with mean zero and
variance $\alpha (1-c)/N$ \cite{DKM89}. There is no contribution
to the learning rule from patterns in the same layer.

The states of the network change as follows. Given a configuration on the
first layer, $\bsigma_N^{1}\equiv \{\sigma_j^{1}\}, \, j=1,...,N$, the state
$\sigma_i^{l+1}$ of unit $i$ on layer $l+1$ is determined exclusively
by the configuration $\bsigma_N^{l}$ of the units in the previous layer
according to the stochastic law
\begin{equation}
\mbox{Prob}(\sigma_i^{l+1}=s\in{ S}|\bsigma_N^{l})
=\frac{\exp[-\beta\epsilon_i(s|\bsigma_N^{l})]}
{\sum_{s\in{S}}\exp[-\beta\epsilon_i(s|\bsigma_N^{l})]} \,\,\,,
\label{3}
\end{equation}
in terms of the single-site energy function on that unit
\begin{equation}
\epsilon_i(s|\bsigma_N^{l})
=-sh_i^{l+1}(\bsigma_N^{l})+\theta_i^{l+1}s^2 \,\,\,,
\label{4}
\end{equation}
where
\begin{equation}
h_i^{l+1}(\bsigma_N^{l})=\sum_{j=1}^NJ_{ij}^{l}\sigma_j^{l},
\label{5}
\end{equation}
is the acting local field and $\theta_i^{l+1}$ is a local
externally adjustable threshold parameter on layer $l+1$. Since
the only changes in the network, in both the synaptic connections
and the states of the units, are between units in consecutive
layers one may associate the layer index with a discrete time step
$t$ and we do this in the following. Thus, the evolution of the
network proceeds according to a parallel dynamics in which the
states of all neurons are updated at each time step.

Next, we consider the relevant quantities that describe the
performance of the network. First, the retrieval overlap between the
state of the network and the pattern $\{\xi_{i}^{\mu,t}\}$ at time
(layer) $t$, given by
\begin{equation}
\tilde{m}_{i}^{\mu,t}={1\over acN}\sum_{j}c_{ij}^t\xi_{j}^{\mu,t}
\sigma_{j}^t \,\,,
\label{6}
\end{equation}
which depends on the site $i$. The dynamical activity and the activity overlap
are defined as
\begin{equation}
q_{0}^t={1\over N}\sum_{i}(\sigma_{i}^t)^{2}\,, \quad
n_{i}^{\mu,t}={1\over
acN}\sum_{i}(c_{ij}^t\xi_{i}^{\mu,t}\sigma_{i}^t)^{2} \,\,,
\label{7}
\end{equation}
respectively. The latter will only be needed for the condensed pattern,
that is the stored pattern to be retrieved.

For the mutual information, we need the conditional probability
distribution $\mbox{Prob}(\sigma_i^{t}|\xi^{\mu,t}_{i})$ that a neuron $i$
is in the state $\sigma_i^{t}$ on layer $t$ given that the $i$-th bit of
the condensed pattern is $\xi^{\mu,t}_{i}$. As a consequence of the
independence of the states of the units on a given layer, it is sufficient
to consider the distribution for a single typical neuron, and we omit here
the index $i$. We also omit here, for clarity, the time index and use
\cite{BD00}
\begin{equation}
\mbox{Prob}(\sigma|\xi^{\mu})=
    (s_{\xi}+m^{\mu}\xi^{\mu}\sigma)\delta(\sigma^{2}-1)+
                           (1-s_{\xi})\delta(\sigma),
    \label{8}
\end{equation}
where
\begin{equation}
      s_{\xi}= s^{\mu}+l^{\mu}(\xi^{\mu})^{2},\quad
      s^{\mu}={q_0-an^{\mu}\over 1-a},\quad
      l^{\mu}={n^{\mu}-q_0\over 1-a}.
 \label{9}
\end{equation}
The mutual information between patterns and neurons, regarding the
patterns as the inputs and the neuron states as the output of the
network channel on each layer, is an architecture independent
property given by \cite{Sh48,Bl90}
\begin{equation}
I^{\mu}(\sigma,\xi^{\mu})
    =S(\sigma)-\overline{S(\sigma|\xi^{\mu})},
\label{10}
\end{equation}
where
\begin{equation}
S(\sigma)=-q_0\ln(q_0/2)-(1-q_0)\ln(1-q_0)
\label{11}
\end{equation}\\
is the entropy and $\overline{S(\sigma|\xi^{\mu})}=
aS_{a}+(1-a)S_{1-a}$
is the  equivocation term with
\begin{eqnarray}
S_{a}&=&-c_{+}^{\mu}\ln c_{+}^{\mu}-c_{-}^{\mu}\ln c_{-}^{\mu}
-(1-n^{\mu})\ln(1-n^{\mu})\nonumber\\
S_{1-a}&=&-s^{\mu}\ln(s^{\mu}/2)-(1-s^{\mu})\ln(1-s^{\mu}).
\label{12}
\end{eqnarray}
Here, $c_{\pm}^{\mu}=(n^{\mu} \pm m^{\mu})/2$ and $s^{\mu}$ is the
parameter in the conditional probability $\mbox{Prob}(\sigma|\xi^{\mu})$.
The mutual information can then be used to obtain the information
content of the network, $i^{\mu}=I^{\mu}\alpha$, where $\alpha=p/cN$ is
the storage ratio.

\section{Recursion Relations}

We consider in this work the retrieval of a single (condensed)
pattern, say $\xi_{i}^{1,t}$, in the dynamics of the network
with a finite overlap $\tilde{m}_{i}^{1,t}=O(1)$ and the remaining
$\mu>1$ overlaps $\tilde{m}_{i}^{\mu,t}=O(1/\sqrt{N})$. The interest
is in the mean overlap,
$m^{t}=[\overline{\langle\tilde{m}_{i}^{1,t}\rangle}]$, where
$[...]$ denotes the average over the $c_{ij}$, $<...>$ denotes the
thermal average with Eq.(3) and the bar means the average over the
patterns. We also need the activity overlap with pattern $\xi_{i}^{1,t}$,
and take $n^{t}=[\overline{\langle n_{i}^{1,t}\rangle}]$.

The recurrence relations that describe the dynamics of the network
for large $N$ follow from the local field, Eq.(5), written as
\begin{equation}
h_i^{t+1}=\xi_{i}^{1,t+1}m^{t}+z\Delta^{t} \,\,\,,
\label{13}
\end{equation}
where the average condensed overlap $m^t=\overline{\langle\sigma^{t}
\rangle \xi^{1,t}/a}$ depends on the local field at time $t-1$ and
$z$ is a Gaussian random variable with zero mean and unit variance
that comes from the action of the macroscopic number of random
overlaps of the diluted network with the uncondensed patterns and
the use of the central limit theorem. The layer-dependent
variance of the local field becomes site independent and is given by
\begin{equation}
(\Delta^{t})^{2}=a\sum_{\mu>1}\overline{[\langle(\tilde{m}_{\mu}^t)^{2}
\rangle]}  \,\,\,.
\label{14}
\end{equation}
A direct calculation in the large-$N$ limit yields
\begin{equation}
(\Delta^{t})^{2}=\alpha (1-c)q_{0}^t+(\Delta_{0}^{t})^{2} \,\,,
\label{15}
\end{equation}
where $\alpha=p/cN$, $q_{0}^t=\overline{\langle(\sigma^t)^{2}
\rangle_{\sigma|\xi}}$ is the dynamical activity and
\begin{equation}
(\Delta_{0}^{t})^{2}=a\sum_{\mu>1}\overline{\langle(m_{\mu}^t)^2
\rangle}
\label{16}
\end{equation}
is now the variance of the Gaussian noise for the connected
layered network in terms of the overlap with the uncondensed
patterns
\begin{equation}
m_{\mu}^t=\frac{1}{aN}\sum_{i}\xi_{i}^{\mu,t}\sigma_{i}^t \,\,.
\label{17}
\end{equation}
Thus, the noise in the local field is a superposition of two Gaussian
noises, one due to the dilution of the synaptic connections and the
other to the macroscopic number of uncondensed patterns, in extension
of an earlier result for the binary network \cite{DKM89}.

Since our main interest in this work is in the effects of synaptic
dilution in a low-activity network, we take a uniform and time-independent
threshold $\theta_{i}^{t}=
\theta$. The averages $\langle\sigma^{t}\rangle$ and
$\langle(\sigma^{t})^{2}\rangle$ are then given, respectively, by
\begin{equation}
F_{\beta}(h^{t},\theta)=
\frac{\sinh(\beta h^{t})}
          {\frac{1}{2}e^{-\beta\theta}+\cosh(\beta h^{t})}\,\,,
\quad
G_{\beta}(h^{t},\theta)=
\frac{\cosh(\beta h^{t})}
     {\frac{1}{2}e^{-\beta\theta}+\cosh(\beta h^{t})}\,\,\,
    \label{18}
\end{equation}
which, in the zero temperature limit, $\beta\rightarrow\infty$,
become
\begin{equation}
F_{\infty}=\mbox{sign}(h^t)\Theta(|h^t|-\theta)\,\,\,,\,\,\,\,\,
G_{\infty}=\Theta(|h^t|-\theta)\,\,\,,
\label{19}
\end{equation}
where $\Theta(x)$ is the usual step function. The performance with
a self-adjusting time-dependent threshold has been considered in a
fully connected layered network \cite{BM00} and in other architectures
[10-14].
Exact dynamic equations are then obtained in the large
$N$ limit in the form of recursion relations for $m^t$, $q_{0}^t$
and $n^t$, where the latter two are needed for the information
content of the network. Similarly, an exact recursion relation for
the second term in the width of the stochastic noise, Eq.(15), is
obtained in the form
\begin{equation}
(\Delta_{0}^{t+1})^{2}=\alpha q_{0}^t+(\chi^{t})^{2}
(\Delta_{0}^{t})^{2} \,\,,
\label{20}
\end{equation}
where
\begin{equation}
\chi^{t}=\beta (q_{0}^{t}-q_{1}^{t})
\label{21}
\end{equation}
is the susceptibility in which $q_{1}^{t}=
\overline{{\langle\sigma^{t}\rangle}^{2}}$. Thus we obtain
\begin{eqnarray}
\label{22}
m^{t+1}&=&\int Dz \,\,
   F_{\beta}(m^{t}+z\Delta^{t}\,,\,\theta)   \\
\label{23}
q_{0}^{t+1}&=&\int Dz \,\,
    \{a\,\,G_{\beta}(m^{t}+z\Delta^{t}\,,\,\theta)
+(1-a)\,\,G_{\beta}(z\Delta^{t}\,,\,\theta)\}  \\
\label{24}
q_{1}^{t+1}&=&\int Dz \,\,
     \{a\,\,F_{\beta}^{2}(m^{t}+z\Delta^{t}\,,\,\theta)
+(1-a)\,\,F_{\beta}^{2}(z\Delta^{t}\,,\,\theta)\} \,\,
\end{eqnarray}
where, as usual, $Dz=\exp(-z^{2}/2)dz/\sqrt{2\pi}$. We also have
\begin{equation}
n^{t+1}=\int Dz \,\,
   G_{\beta}(m^{t}+z\Delta^{t}\,,\,\theta) \,\,.
\label{25}
\end{equation}

The dynamics, including transients, and the stationary states of the
diluted layered network follows then from the solutions of these
equations together with Eq.(20). The stationary states are reached
when $m^{t+1}=m^{t}$, $q_{0}^{t+1}=q_{0}^{t}$, $q_{1}^{t+1}=q_{1}^{t}$,
$n^{t+1}=n^{t}$ and $\Delta_{0}^{t+1}=\Delta_{0}^{t}$, and we call the
first three, respectively, $m$, $q_{0}$ and $q_1$. The phase diagrams for
the stationary states and the time evolution of the order parameters are
discussed in the next section.

\section{Dynamics and stationary states}

The stable stationary states yield one or more retrieval phases $R(m>0,
q_1>0)$ and a spin-glass phase $SG(m=0,q_1>0)$, all as sustained activity
solutions with $q_{0}>0$. Since the model is a dynamical one, the stability
criterium is that the change in the order parameters should become smaller
in every one of the final steps of the iteration procedure of the flow
towards an attractor fixed point. Thus, for the retrieval overlap one has
\begin{equation}
\lim_{{\delta} m(t)\rightarrow 0}\frac{{\delta} m(t+1)}
{{\delta} m(t)}<1 \,\,\,,
\label{26}
\end{equation}
and similar relationships for the other parameters.

We are interested in this paper in the characteristic features of
finite dilution of the phase diagrams and in the specific
performance of the network. Different kinds of phase diagrams are
obtained depending on the pattern activity $a$ and on $c$. In the
case of full connectivity ($c=1$) and low $a$, we find the
$(\alpha,\theta)$ phase diagram shown in Fig. 1 for $a=0.5$
and either $T=0$ (full lines) or $T=0.05$ (dashed lines). The
lines represent discontinuous phase boundaries where the locally
stable retrieval states appear with decreasing load $\alpha$ below
a critical $\alpha_c$ and the dotted line indicates the locus of
optimal performance which yields the largest load capacity that
still sustains retrieval behavior. There is a weak retrieval phase
(I) and a strong retrieval phase (II) separated by a discontinuous
phase boundary. There is also a smaller retrieval phase in the lower
triangular region. For larger activity and full connectivity, as
in the case of uniform patterns ($a=2/3$), there is mostly a single
retrieval phase with a strong $\alpha$ dependent optimal performance
and this feature remains even for finite dilution, say for $c=0.5$.

\begin{figure}
\vspace*{1cm}
\centerline{
\includegraphics[width=.46\textwidth,height=10cm, angle=270]{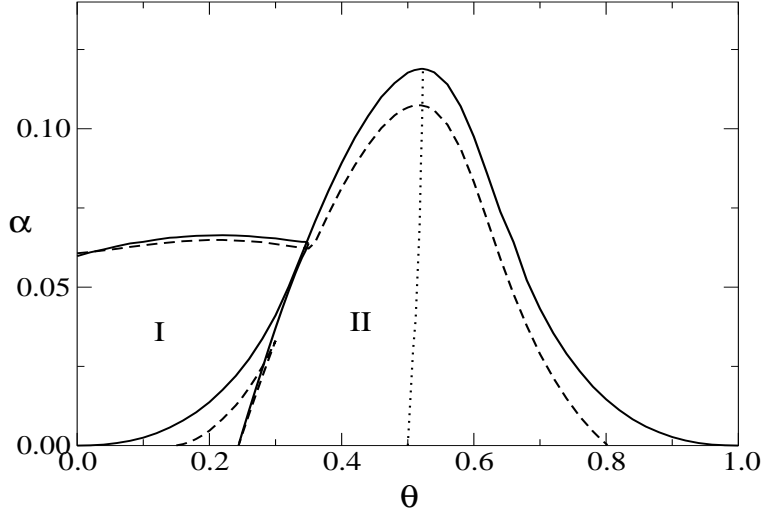}
}
\hfill
\caption{Storage capacity vs. threshold ($\alpha,\theta$) phase diagram for
the fully connected $Q=3$ Ising layered feed-forward network, with activity
$a=0.5$, $T=0$ (full lines) and $T=0.05$ (dashed lines). The phases are
described in the text and the dotted line is the locus of optimal performance.}
\end{figure}

A different situation appears for both finite dilution, with $c$
smaller than a critical activity-dependent $c^{*}$, and for low
activity where part of the phase boundary between regions I and II
disappears. For $a=0.5$ this is the case when $c^{*}=0.87$ at
$\alpha=0.34$ and $T=0$. As shown in Fig. 2, for typical $c=0.8$
and $a=0.5$, there is now a gradual transition from region I to
region II and, as $c$ decreases, the continuous transition region
increases until a stage is reached where the maxima in both
regions merge into a single maximum with a lower optimal threshold
$\theta$. The phase diagram in Fig. 2 is reminiscent of the
diagram for the finite diluted recurrent $Q=3$ state network with
uniform patterns \cite{TE01}.

\begin{figure}
\vspace*{1cm}
\centerline{
\includegraphics[width=.46\textwidth,height=10cm, angle=270]{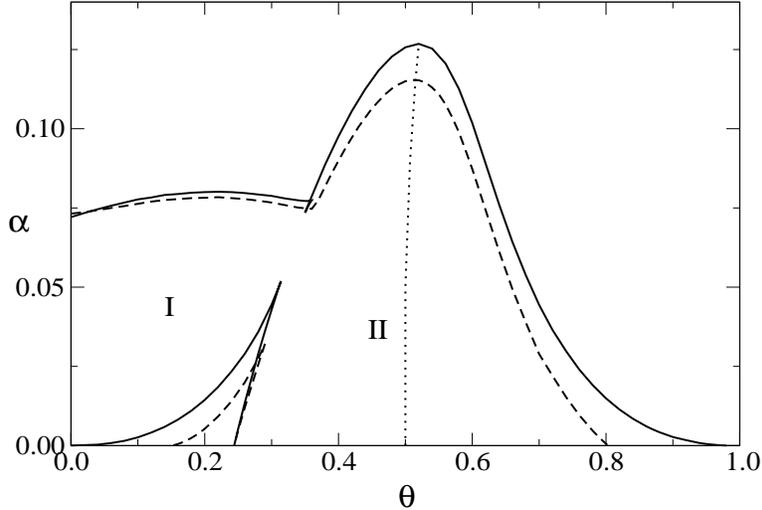}
}
\hfill
\caption{($\alpha,\theta$) phase diagram for the diluted network, with $c=0.8$,
$a=0.5$, for $T=0$ (full lines) and $T=0.05$ (dashed lines). The phases are
described in the text.}
\end{figure}

In order to study the dynamics and the stability of the phases, we
consider the time evolution of the retrieval overlap shown in Fig.
3 for $c=0.8$, $a=0.5$, $\alpha=0.06$, $T=0$ and $\theta$ between
$0.293$ and $0.33$. For small $\theta$ the retrieval state is not
stable as shown by the typical lower curve ($0.293$), and there is
a whole part of region I where this is the case. Since it is not
clear that these instabilities have any significance, we do not
explore this issue further in this work. On the other hand, as
$\theta$ increases, the retrieval state becomes stable and a
maximum overlap is reached after a relatively short transient.

\begin{figure}
\vspace*{1cm}
\centerline{
\includegraphics[width=.46\textwidth,height=10cm, angle=270]{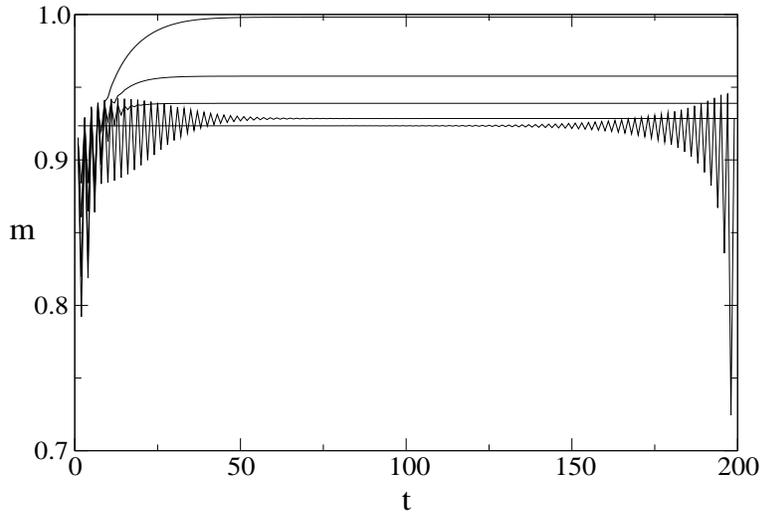}
}
\hfill
\caption{Time evolution of the retrieval overlap $m$ for the diluted network 
with $c=0.8$, $a=0.5$, $T=0$ and $\alpha=0.06$, for $\theta=0.293$, $0.30$, 
$0.31$, $0.32$ and $0.33$, from bottom to top. Note the instability for large 
$t$ in the first case, and the transients for the other cases.}
\end{figure}

A further property of the network is the increase in the maximum
information content with dilution as shown near to the optimal
performance with $\theta=0.5$ for $a=0.5$ and various values of
$c$ in Fig. 4. Note that the result of a non-zero information
content for increasing $\alpha$ with dilution is not surprising
since we defined $\alpha=p/cN$, but the increase of the maximum
information content with dilution is a novel feature.

\begin{figure}
\vspace*{1cm}
\centerline{
\includegraphics[width=.46\textwidth,height=10cm, angle=270]{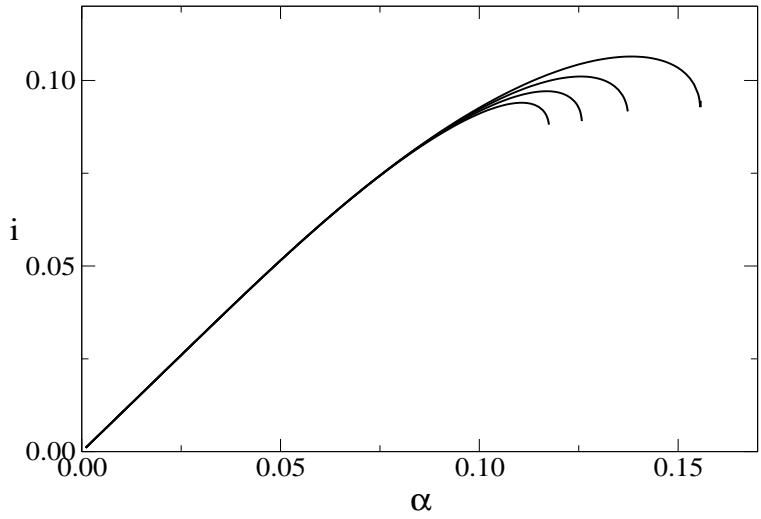}
}
\hfill
\caption{Information content $i=I\alpha$ for $a=0.5$, $\theta=0.5$ 
and $c=1$ (fully connected layered network), $c=0.8$, $0.6$ and $0.4$, from
bottom to top.}
\end{figure}

There is also a considerable increase in the maximum information
content with an increasing threshold in the good performance
region II, as shown in Fig. 5 for $c=0.8$, $a=0.5$ and various
values of $\theta$. Also, the decrease from the maximum is
smoother with dilution than in the case of the fully connected
layered network. Finally, one may consider the robustness of the
network to synaptic noise and in Fig. 6 we show the maximum
information content, $i_{max}$ for $a=0.5$, $\theta=0.5$ and
various values of $c$. Clearly, there is a more gradual decrease
in performance with $T$ for lower connectivity.

\begin{figure}
\vspace*{1cm}
\centerline{
\includegraphics[width=.46\textwidth,height=10cm, angle=270]{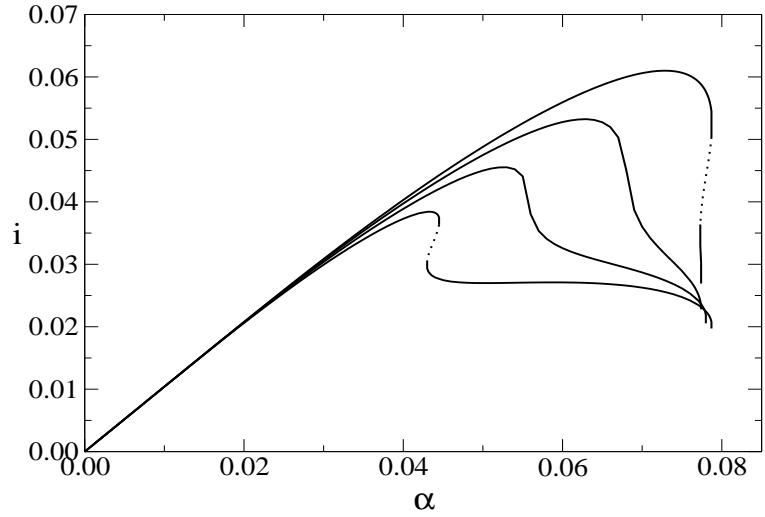}
}
\hfill
\caption{Information content $i$ for $a=0.5$, $c=0.8$ and threshold 
$\theta=0.30$, $0.32$, $0.34$ and $0.36$, from bottom to top. The dotted parts 
of the lines indicate unstable states.}
\end{figure}

\begin{figure}
\vspace*{1cm}
\centerline{
\includegraphics[width=.46\textwidth,height=10cm, angle=270]{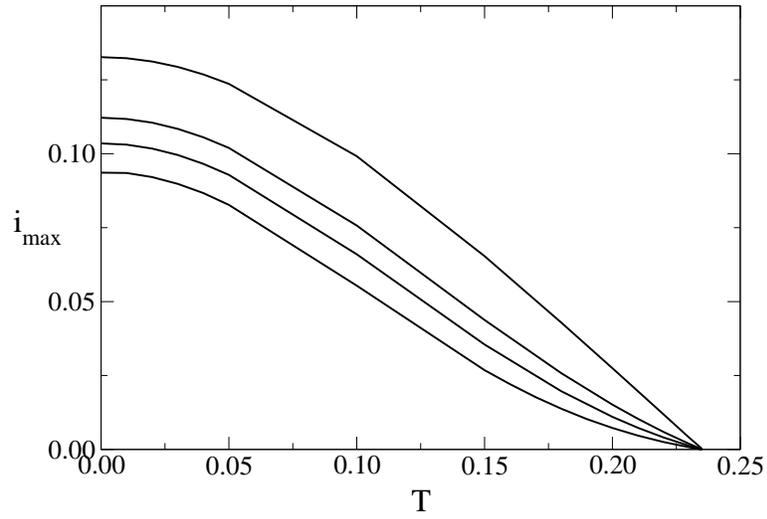}
}
\hfill
\caption{Maximum information content $i_{max}$ as a function of the 
synaptic noise $T$ for $a=0.5$, $\theta=0.5$ and $c=1$ (fully connected
layered network), $0.5$, $0.25$ and $0$ (extremely diluted network, from
bottom to top.}
\end{figure}

\section{Summary and conclusions}

We derived the exact recursion relations in the large-$N$ limit
that describe the time evolution and the phase diagrams for the
stationary states of the macroscopic variables in a three-state
layered feed-forward network with finite synaptic dilution.
Synaptic dilution appears as a static stochastic noise for a
macroscopic number of stored patterns \cite{DKM89,S86}. This is a
model with asymmetric interactions between units in consecutive
layers and we allow for variable pattern activity in training the
network with ternary patterns. Instead of discontinuous phase
boundaries between retrieval and non-retrieval states, or between
qualitatively different retrieval states for full synaptic
connectivity, we find that there can be a continuous change with
no boundary at all from weak to optimal retrieval states with a
varying threshold in the local field, for low pattern activity.
Phase boundaries of continuous transitions are known to appear in
extremely diluted symmetric or asymmetric networks, but we emphasize
that the continuous changeover in the present model is due to the
joint action of {\it finite} dilution and low activity patterns.

In view of a similar recent result for a three-state recurrent diluted
network with symmetric synaptic connections \cite{TE01}, one may conclude
that this is a feature of finite dilution which is independent of both
the network architecture and of the interaction symmetry. Thus, provided
there is an above minimum threshold such that a network attains the
ability to retrieve a nominated pattern after eliminating undesirable
transient states, the good retrieval performance does not depend in an
essential way on a precise threshold adjustment and this may explain why
biological networks, in which there are no precise thresholds, can have
a good performance despite a fraction of missing synaptic connections.
This is an activity dependent property and it might help in the study of
plasticity in neural networks. It may also be useful for artificial
neural networks.

It may be worthwhile to note the asymmetric dual role of the
threshold dependence discussed in this work. Whereas there is a
continuous improvement in network performance for low to moderate
threshold, there is an abrupt end to the performance for large
threshold, as one would expect, since in the latter case mostly
the inactive states of the network become dominant. We also showed that 
there is an improvement with finite dilution in both the size of the 
information content transmitted by the network and in the
continuous changes of the information with a varying threshold. That is,
as long as there is a convergence to stable stationary states, here again
the network performance seems to be less sensitive to threshold adjustment.

The network behavior discussed in this work is also expected to
appear in a diluted layered $Q=4$ state network with low activity
patterns, based on recent results on a recurrent network which
exhibits a continuous improvement in network performance with
varying low-to-moderate threshold for low activity patterns \cite{TE01}.

\section{Acknowledgments}

The work of one of the authors (WKT) was financially supported, 
in part, by CNPq (Conselho Nacional de Desenvolvimento Cient\'{\i}fico 
e Tecnol\'ogico), Brazil.


\begin{thebibliography}{99}

\bibitem{RTFP97} E.T. Rolls, A. Treves, D. Foster and C. Perez-Vicente,
Neural Networks {\bf 10}, 1559 (1995).
\bibitem{Ar95} {\it The Handbook of Brain Theory and Neural Networks},
M.A. Arbib, editor (M.I.T. Press, 1995).
\bibitem{TE01} W. K. Theumann and R. Erichsen Jr. Phys. Rev. E {\bf 64},
061902 (2001).
\bibitem{Co01} A. C. Coolen, {\it Handbook of Biological Physics IV}, 597,
(F. Moss and S. Gielen (Eds.), Elsevier Science, Amsterdam 2001)
\bibitem{DKM89} E. Domany, W. Kinzel and R. Meir, J. Phys. A {\bf 22},
2081(1989).
\bibitem{BSV94} D. Boll\'e, G. M. Shim and B. Vinck, J. Stat. Phys.
{\bf 74}, 583 (1994).
\bibitem{BD00} D. Boll\'e and D. Dominguez Carreta, Physica A {\bf 286},
401(2000).
\bibitem{Sh48} C. E. Shannon, Bell Syst. Techn. J. {\bf 27}, 379 (1948).
\bibitem{Bl90} R. E. Blahut, {\it Principles and Practice of Information
Theory}(Addison-Wesley, Reading, MA 1990), Chapter 5.
\bibitem{BM00} D. Boll\'e and G. Massolo, J. Phys. A {\bf 33}, 2597
(2000).
\bibitem{DB98} D. R. C. Dominguez and D. Boll\'e, Phys. Rev. Lett.
{\bf 80}, 2961 (1998).
\bibitem{KA98} K. Kitano and T. Aoyagi, J. Phys. A {\bf 31}, L613 (1998).
\bibitem{SG99} S. Grosskinsky, J. Phys. A {\bf 32}, 22983 (1999).
\bibitem{BDA00} D. Boll\'e, D. R. C. Dominguez and S. Amari, Neural
Networks {\bf 13}, 455 (2000).
\bibitem{S86} H. Sompolinsky, Phys. Rev. {\bf 34 },2571 (1986).

\end{thebibliography}
\end{document}